\documentclass[final,3p,11pt,authoryear]{elsarticle}

\makeatletter
\def\ps@pprintTitle{%
 \let\@oddhead\@empty
 \let\@evenhead\@empty
 \def\@oddfoot{\centerline{\thepage}}%
 \let\@evenfoot\@oddfoot}
\makeatother
\usepackage{color}
\usepackage[usenames,dvipsnames]{xcolor}
\usepackage{hyperref}

\usepackage{natbib}
\biboptions{numbers,sort&compress}
\usepackage[nodots]{numcompress}
\usepackage{graphicx}
\usepackage{caption}
\usepackage{subcaption}
\usepackage{amsfonts,amssymb,amsbsy,amsmath,mathtools}

\journal{Elsevier}
\begin{document}

\begin{frontmatter}

\renewcommand{\thefootnote}{\fnsymbol{footnote}}
\title{\textbf{Nonlinear finite element analysis of lattice core sandwich beams\let\thefootnote\relax\footnote{{\color{Blue}\textbf{Recompiled, unedited accepted manuscript}}. \copyright 2019. Made available under \href{https://creativecommons.org/licenses/by-nc-nd/4.0/}{{\color{Blue}\textbf{\underline{CC-BY-NC-ND 4.0}}}}}}}

%% Title, authors and addresses

%% use the tnoteref command within \title for footnotes;
%% use the tnotetext command for theassociated footnote;
%% use the fnref command within \author or \address for footnotes;
%% use the fntext command for theassociated footnote;
%% use the corref command within \author for corresponding author footnotes;
%% use the cortext command for theassociated footnote;
%% use the ead command for the email address,
%% and the form \ead[url] for the home page:
%% \title{Title\tnoteref{label1}}
%% \tnotetext[label1]{}
%% \author{Name\corref{cor1}\fnref{label2}}
%% \ead{email address}
%% \ead[url]{home page}
%% \fntext[label2]{}
%% \cortext[cor1]{}
%% \address{Address\fnref{label3}}
%% \fntext[label3]{}

%% use optional labels to link authors explicitly to addresses:
%% \author[label1,label2]{}
%% \address[label1]{}
%% \address[label2]{}

\author[add2]{Praneeth Nampally}
\address[add2]{Texas A\&M University, Department of Mechanical Engineering, College Station, TX 77843-3123, USA}
\author[add2,add1]{Anssi T. Karttunen\corref{cor1}}
\cortext[cor1]{Corresponding author.  anssi.karttunen@iki.fi.\textbf{Cite as}: \textit{Eur. J. Mech. A-Solid} 2019;160:66--75 \href{https://doi.org/10.1016/j.euromechsol.2018.12.006}{{\color{OliveGreen}\textbf{\underline{doi link}}}}}
\address[add1]{Aalto University, Department of Mechanical Engineering, FI-00076 Aalto, Finland}
\author[add2]{J.N. Reddy}

\begin{abstract}
A geometrically nonlinear finite element model is developed for the bending analysis of micropolar Timoshenko beams using the principle of virtual displacements and linear Lagrange interpolation functions.  The nonlinearity enters the model via a nonlinear von K\'arm\'an strain term that allows the micropolar beam to undergo moderate rotations. The nonlinear micropolar Timoshenko beam is used as an equivalent single layer model to study four different lattice core sandwich beams. A two-scale energy method is used to derive the micropolar constitutive equations for web, hexagonal, Y-frame and corrugated core topologies. Various bending cases are studied numerically using the developed 1-D finite element model. Reduced integration techniques are used to overcome the shear and membrane locking. The present 1-D results are in good agreement with the corresponding 2-D finite element beam frame results for global bending.
\end{abstract}
\begin{keyword}
%% keywords here, in the form: keyword \sep keyword
Micropolar beam \sep Constitutive modeling \sep Geometric nonlinearity \sep Lattice material \sep Finite element \sep Nonlinear bending
%% PACS codes here, in the form: \PACS code \sep code

%% MSC codes here, in the form: \MSC code \sep code
%% or \MSC[2008] code \sep code (2000 is the default)

\end{keyword}

\end{frontmatter}

%% \linenumbers

%% main text
\section{Introduction}
A typical sandwich panel consists of a thick, low-stiffness core between two relatively thin but stiff face sheets. The face sheets take bending and in-plane loads while the core carries transverse shear loads \citep{allen1969,vinson1999}. The face sheets and core can be made of the same or different materials and some possible core structures include, for example, foam, solid, honeycomb, and truss cores \citep{vinson2001}. A number of manufacturing techniques are available for sandwich panels \citep{Karlsson1997,wadley2003}. A sandwich panel generally has a high bending stiffness compared to a single solid plate of the same dimensions made of either the face sheet or core material and the panel weighs considerably less than the solid plate making it a weight-efficient structure \citep{vinson1999}. Most early sandwich panels had isotropic face sheets but with the development of fiber reinforced composites, construction of sandwich panels with composite face sheets has become possible. Because of their high stiffness-to-weight ratios, structural efficiency and durability, sandwich panels are suitable candidates for high-speed aircraft and space applications \citep{Schwingel2007}. They have also found applications, for example, in shipbuilding \citep{roland1997,kujala2005} and other marine applications \citep{bitzer1994,Mouritz2001,knox1998} and in the construction of bridges and buildings \citep{Davalos2001,bright2004,bright2007,nilsson2017,briscoe2011}.

The number of applications for sandwich panels is increasing rapidly. The required accuracy in the structural analysis of the panels depends on the type of the application considered. For example, in air-crafts a very detailed response of the sandwich structure may be required, whereas an overall global response may suffice in residential buildings when the natural vibration frequencies are of interest, for example. In any case, there is a need for appropriate modeling tools for different applications. Reviews on the modeling of sandwich structures have been given by several authors \citep{burton1995,hohe2004,carrera2009,sayyad2017,birman2018}. Modeling methods for sandwich panels can be broadly classified as: (a) Complete 3-D analysis (computational or analytical), with complete details of the face sheets and the core structure considered; (b) layer-wise modeling with the faces and core considered as separate continuum layers \citep{reddy2004}; (c) statically equivalent single layer (ESL) models. Although computational 3-D and layer-wise analyses give very detailed stress distributions for the panels, they come with the inherent disadvantage of including a large number of variables and, thus, the computational analysis of them can be very burdensome. Therefore, equivalent single layer theories such as the ESL first-order shear deformation (FSDT) beam and plate models are attractive especially when the global response of the structure is of main interest without accounting for every small detail. Extensive literature exists on the modeling of sandwich beam, plates and shells by ESL theories, see, for example \citep{Skvotsov2001,Barut2001,Barut2002,hohe2004,abrate2017}.

It was shown recently that all-steel web-core sandwich beams deform so that when an ESL-FSDT model is used, the model has to take into account anti-symmetric shear deformations in order for the response of the sandwich structure to be captured accurately in some applications \citep{karttunen2018a}. In more detail, the constituents of a 2-D or 3-D web-core sandwich beam model do not exhibit any anti-symmetric shear strains, but when the problem is reduced to a 1-D ESL beam problem essentially by thickness integration, the anti-symmetric behavior needs to be considered via a 1-D micropolar Timoshenko beam model. In this paper, we develop a geometrically nonlinear finite element model based on the micropolar Timoshenko beam theory.

With the revived interest in micropolar elasticity \citep{eringen1964}, considerable work has been put into developing appropriate finite element models for micropolar continua in general; see, for example, \citep{pothier1994,Li2004,roman2013,Zhou2015}. To list a few recent finite element models for micropolar plates we mention the works of \cite{ansari2016,ansari2018} and \cite{godio2014}. Various finite element models have been proposed for the bending analysis of micropolar beams as well. \cite{huang2000} used 3-D non-compatible finite elements to analyze the bending of beams, and \cite{Li2004} proposed three different elements for plane micropolar elasticity and used them to analyze thin in-plane beams. \cite{hassanpour2014} developed a 1-D micropolar beam finite element model using Lagrange interpolation functions. \cite{regueiro2015} derived a finite element model for a micropolar Timoshenko beam with the microrotation assumed to be equal to the cross-sectional rotation. More recently, \cite{karttunen2018a} proposed nodally-exact 1-D finite element to analyze micropolar Timoshenko beams and \cite{ansari2018} proposed a 27-node 3-D finite element for the analysis of beams. Only linear strains were considered in developing the finite element models in all the above mentioned papers.

In this study, in order to develop the geometrically nonlinear micropolar 1-D beam finite element, we start from the principle of virtual displacements and use the linear Lagrange interpolation functions. The originally linear micropolar Timoshenko beam model of \cite{karttunen2018a} is enriched by nonlinear von K\'arm\'an strains in order to account for the moderate rotations of the beam \citep{ding2016}. Lattice cores, namely, hexagonal, corrugated and Y-frame cores in addition to the web-core topology studied earlier are considered. To this end, the two-scale constitutive modeling method presented by \cite{karttunen2018b} is first generalized to cores other than the web-core.

In more detail, the rest of the paper is organized as follows. In Section 2, a brief review on the linear 1-D micropolar Timoshenko beam model is given followed by the two-scale constitutive modeling of the four different lattice cores after which the  geometrically nonlinear micropolar Timoshenko beam equations are derived. In Section 3, the displacement based finite element model is formulated using the principle of virtual displacements. The element stiffness matrices are derived and the used nonlinear iterative procedures and techniques to avoid numerical locking are discussed. In Section 4, numerical examples are solved using the developed 1-D micropolar beam finite element model and are compared with the corresponding results from 2-D finite element analyses. Finally, concluding remarks are given in Section 5.

\section{Geometrically nonlinear micropolar Timoshenko beam}
Two-scale constitutive modeling of lattice materials in the context of linear micropolar Timoshenko beam theory is first carried out. The strain energy density for a geometrically nonlinear micropolar Timoshenko beam is obtained by retaining the linear constitutive matrix while the strain vector is augmented with von K\'arm\'an nonlinearity. The hyperelastic constitutive relations, i.e., the stress resultant equations for the nonlinear beam are derived from the strain energy for four lattice materials. The equilibrium equations for the nonlinear beam in terms of the stress resultants, and the corresponding weak form for finite element developments are attained by employing the principle of virtual displacements.
\subsection{Displacements and linear strains}
The two-dimensional displacements $U_x$ and $U_y$ and the independent microrotation $\Psi$ of a micropolar Timoshenko beam can be written in terms of central axis kinematic variables as \citep{karttunen2018b}
\begin{equation}
U_x(x,y)=u_x(x)+y\phi(x), \quad U_y(x,y)=u_y(x), \quad \Psi(x,y)=\psi(x),
\end{equation}
where $u_x$ is the axial displacement, $\phi$ is the rotation of the cross-section, $u_y$ is the transverse deflection, and $\psi$ is the microrotation. The nonzero infinitesimal strains of the beam are
\begin{equation}
\begin{aligned}
\varepsilon_x&=\frac{\partial U_x}{\partial x}=u_x'+y\phi'=\varepsilon_x^0+y\kappa_{x}, \ & \kappa_{xz}&=\frac{\partial \Psi}{\partial x}=\psi' \\ \varepsilon_{xy}&=\frac{\partial U_y}{\partial x}-\Psi=u_y'-\psi, \ & \varepsilon_{yx}&=\frac{\partial U_x}{\partial y}+\Psi=\phi+\psi,
\end{aligned}
\end{equation}
where the prime ``$'$" on the variables denotes differentiation with respect to $x$. The symmetric and anti-symmetric shear strains of the beam are defined as
\begin{align}
\gamma_s&=\varepsilon_{xy}+\varepsilon_{yx}=u_y'+\phi, \\ \gamma_a&=\varepsilon_{xy}-\varepsilon_{yx}=u_y'-\phi-2\psi=2(\omega_z-\psi),
\end{align}
respectively. The anti-symmetric part is defined by the difference between the macrorotation $\omega_z$ and the microrotation $\psi$. For $\omega_z=\psi$, the relative strains reduce to their classical definitions \citep{barber2010}, for example, $\varepsilon_{xy}=u_y'-\omega_z=(u_y'+\phi)/2$.
\subsection{Two-scale constitutive modeling}
Figure 1 shows a rectangular unit cell attached to an arbitrary cross section of a micropolar Timoshenko beam. The unit cell of length $l$ and height $h$ represent the periodic microstructure of the macrostructural beam ($l\leq\textrm{beam length}$). The unit cell corner displacements in Fig.~1 are expressed in terms of the micropolar cross-sectional displacements $U_x$ and $U_y$ and rotation $\Psi$. With distance from an arbitrary beam cross section, Taylor series expansions of Eqs.~(1) lead to
\begin{figure}
\centering
\includegraphics[scale=1.1]{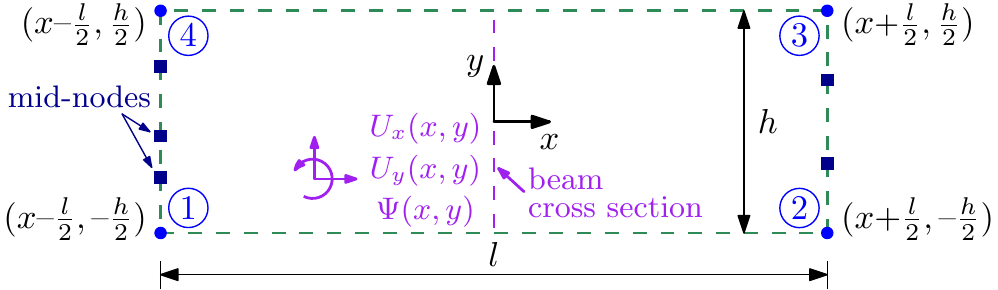}
\caption{Arbitrary cross section of the micropolar beam possessing microstructure of length $l$ $(\textrm{beam length}\geq l$).}
\end{figure}
\begin{align}
U_x(x\pm l/2,\pm h/2)&= u_x\pm\frac{h}{2}\left[\frac{1}{2}(\gamma_s-\gamma_a)-\psi\right]\pm\frac{l}{2}\left(\varepsilon_x^0\pm\frac{h}{2}\kappa_x\right) , \\
U_y(x\pm l/2,\pm h/2)&=u_y\pm\frac{l}{2}\left[\frac{1}{2}(\gamma_s+\gamma_a)+\psi\right] , \\
\Psi(x\pm l/2,\pm h/2)&=\psi\pm\frac{l}{2}\kappa_{xz} ,
\end{align}
where the micropolar strains (2)--(4) have been imposed on the cross-sectional rotation $\phi$ and the displacement gradients. Using the node numbering of Fig.~1, we can write the discrete-to-continuum transformation given by Eqs.~(5)--(7) for the corner nodes in matrix form
\begin{equation}
\mathbf{d}^c=\mathbf{T}^c_u\mathbf{u}+\mathbf{T}^c_\varepsilon\boldsymbol{\varepsilon},
\end{equation}
where the generalized discrete displacement vector is
\begin{equation}
\mathbf{d}^c=\left\{U_{x,1} \ \ U_{y,1} \ \ \Psi_{1} \ \ U_{x,2} \ \ U_{y,2} \ \ \Psi_{2} \ \ U_{x,3} \ \ U_{y,3} \ \ \Psi_{3} \ \ U_{x,4} \ \ U_{y,4} \ \ \Psi_{4}\right\}^{\textrm{T}}
\end{equation}
and the vectors for the continuous variables read
\begin{align}
\mathbf{u}&=\left\{u_x \ \ u_y \ \ \phi \ \ \psi\right\}^{\textrm{T}} , \\
{\boldsymbol\varepsilon}&=\{\varepsilon_x^0 \quad \kappa_x \quad \gamma_s \quad \gamma_a \quad \kappa_{xz}\}^{\textnormal{T}} .
\end{align}
The transformation matrices $\mathbf{T}^c_u$ and $\mathbf{T}^c_\varepsilon$ are given in Appendix A. For additional mid-nodes located at $\pm l/2$ and connected to the neighboring unit cells, a similar transformation can be written as
\begin{equation}
\mathbf{d}^m=\mathbf{T}^m_u\mathbf{u}+\mathbf{T}^m_\varepsilon{\boldsymbol \varepsilon}.
\end{equation}
The complete discrete-to-continuum transformation reads
\begin{equation}
\mathbf{d}=
\begin{Bmatrix}
\mathbf{d}^c \\
\mathbf{d}^m
\end{Bmatrix}
=
\begin{bmatrix}
\mathbf{T}^c_u & \mathbf{T}^c_\varepsilon \\
\mathbf{T}^m_u & \mathbf{T}^m_\varepsilon
\end{bmatrix}
\begin{Bmatrix}
\mathbf{u} \\
{\boldsymbol \varepsilon}
\end{Bmatrix}.
\end{equation}
The transformation by Eq.~(13) for nodes located at the unit cell edges $x=\pm l/2$ may be applied to different finite element based lattice unit cells once static condensation has been applied at all nodes located between $l/2<x<l/2$. The unit cell can be modeled, for example, by using Euler-Bernoulli or Timoshenko beam finite elements as both choices result in a system that is consistent with the generalized displacement vector (13). As for their material composition, the elements can be heterogeneous and anisotropic. Several lattice cores will be considered in Section 2.3.

The strain energy of a unit cell modeled by beam elements can be written as
\begin{equation}
W=\frac{1}{2}\mathbf{d}^{\textrm{T}}\mathbf{k}\mathbf{d},
\end{equation}
where $\mathbf{k}$ is the global stiffness matrix of the unit cell that corresponds to the master degrees of freedom after the static condensation. It is straightforward to verify that the displacement terms (10) do not contribute to the strain energy of any of the lattice core unit cells considered in this paper and we obtain
\begin{equation}
W=\frac{1}{2}{\boldsymbol\varepsilon}^{\textrm{T}}\mathbf{T}_\varepsilon^{\textrm{T}}\mathbf{k}\mathbf{T}^{\phantom{ }}_\varepsilon{\boldsymbol \varepsilon}.
\end{equation}
We define the 1-D density of the unit cell strain energy as
\begin{equation}
W_0^l\equiv\frac{W}{l}=\frac{1}{2}{\boldsymbol \varepsilon}^{\textrm{T}}\mathbf{C}{\boldsymbol \varepsilon}
\end{equation}
where the constitutive matrix is given by
\begin{equation}
\mathbf{C}=\frac{1}{l}\mathbf{T}_\varepsilon^{\textrm{T}}\mathbf{k}\mathbf{T}^{\phantom{ }}_\varepsilon.
\end{equation}
For geometrically nonlinear unit cell analysis in which strains are small but rotations may be moderate, we assume that the constitutive matrix (17) remains the same but the axial normal strain is modified so that the strain vector becomes
\begin{equation}
{\hat{\boldsymbol\varepsilon}}=\{\hat{\varepsilon}_x^0 \quad \kappa_x \quad \gamma_s \quad \gamma_a \quad \kappa_{xz}\}^{\textnormal{T}} .
\end{equation}
in which the nonlinear von K\'arm\'an term is included in
\begin{equation}
\hat{\varepsilon}_x^0=u_x'+\frac{1}{2}\left(u_y'\right)^2.
\end{equation}
The unit cell represents a \textit{lattice material} of which the micropolar beam is made of. We write the hyperelastic constitutive relations for the geometrically nonlinear micropolar beam continuum as
\begin{equation}
\mathbf{S}\equiv\frac{\partial \hat{W}_0^l}{\partial {\hat{\boldsymbol\varepsilon}}}=\frac{1}{2}\frac{\partial }{\partial {\hat{\boldsymbol\varepsilon}}}\left({\hat{\boldsymbol\varepsilon}}^{\textrm{T}}\mathbf{C}{\hat{\boldsymbol\varepsilon}}\right)=\mathbf{C}\boldsymbol{\hat{\varepsilon}},
\end{equation}
where $\mathbf{S}$ is now the stress resultant vector of the 1-D micropolar beam and for which the general explicit form considered in this paper is
\begin{equation}
\begin{Bmatrix}
N_x \\
M_x \\
Q_s \\
Q_a \\
P_{xz}
\end{Bmatrix}
=
\begin{bmatrix}
C_{11} & C_{12} & 0 & 0 & C_{15} \\
C_{12} & C_{22} & 0 & 0 & C_{25} \\
0 & 0 & C_{33} & C_{34} & 0 \\
0 & 0 & C_{34} & C_{44} & 0 \\
C_{15} & C_{25} & 0 & 0 & C_{55}
\end{bmatrix}
\begin{Bmatrix}
\hat{\varepsilon}_x^0 \\
\kappa_x \\
\gamma_s \\
\gamma_a \\
\kappa_{xz}
\end{Bmatrix},
\end{equation}
where $N_x$ is the axial force, $M_x$ and $P_{xz}$ are the bending and couple-stress moments, respectively, and $Q_s$ and $Q_a$ are the symmetric and anti-symmetric shear forces defined as \cite{karttunen2018b}
\begin{equation}
Q_s=\frac{Q_{xy}+Q_{yx}}{2} \quad \textrm{and} \quad
Q_a=\frac{Q_{xy}-Q_{yx}}{2},
\end{equation}
respectively. Note that Eq.~(20) implies that the bridging of the two scales, i.e., the macroscale (beam) and the microscale (unit cell), is founded on an assumption of strain energy equivalence between them.
\subsection{Modeling of lattice materials}
In this section, we consider four different lattice material sandwich beam cores, namely, the web-core studied earlier in the linear context \citep{karttunen2018b} and the hexagonal, Y-frame and corrugated cores presented in Fig.~2. All cores are made of steel and are modeled using linearly elastic isotropic nodally-exact Euler-Bernoulli beam finite elements. All unit cell beam constituents have a rectangular cross-section and in this paper the width of all unit cells is 0.05 m.

The web-core is modeled using four Euler-Bernoulli beam elements that can represented by the dashed lines in Fig.~1. In this case, only the corner node transformation (8) needs to be considered and the application of this in Eq.~(14) results in the constitutive equations
\begin{equation}
\begin{Bmatrix}
N_x \\
M_x \\
Q_s \\
Q_a \\
P_{xz}
\end{Bmatrix}
=
\begin{bmatrix}
 2 EA_f & 0 & 0 & 0 & 0 \\
  & \frac{EA_f h^2}{2}+\Theta  & 0 & 0 & \Theta  \\
  &  & \frac{6EI_f+\Theta}{l^2} & \frac{6EI_f-\Theta}{l^2} & 0 \\
  & \textrm{SYM}  &  & \frac{6EI_f+\Theta}{l^2} & 0 \\
  & &  &  & 2EI_f +\Theta \\
\end{bmatrix}
\begin{Bmatrix}
\hat{\varepsilon}_x^0 \\
\kappa_x \\
\gamma_s \\
\gamma_a \\
\kappa_{xz}
\end{Bmatrix}
\end{equation}
where
\begin{equation}
\Theta=\frac{3EI_w k_\theta l}{6EI_w+k_\theta h}.
\end{equation}
In the constitutive matrix, $EA_f$ and $EI_f$ are the axial and bending stiffnesses of the horizontal faces, respectively. For the vertical webs we have $EA_w$, $EI_w$ and $k_\theta$ for the axial, bending and rotational joint stiffnesses, respectively. The webs are modeled using special-purpose Euler-Bernoulli elements with rotational springs at both ends to account for the flexibility of the laser-welded joints \citep{monforton1963,chen2005,romanoff2007c}. The numerical values of the core parameters are  $E_f=212$ GPa, $E_w=200$ GPa, $k_\theta=2675$ Nm and $\nu=0.3$ for the face and web Young's moduli, rotational joint stiffness and Poisson ratio, respectively. The face and web thicknesses are $t_f=3$ mm and $t_w=4$ mm, respectively. The height, i.e, the distance between the face central axes is $h=43$ mm.
\begin{figure}
\centering
\includegraphics[scale=1.0]{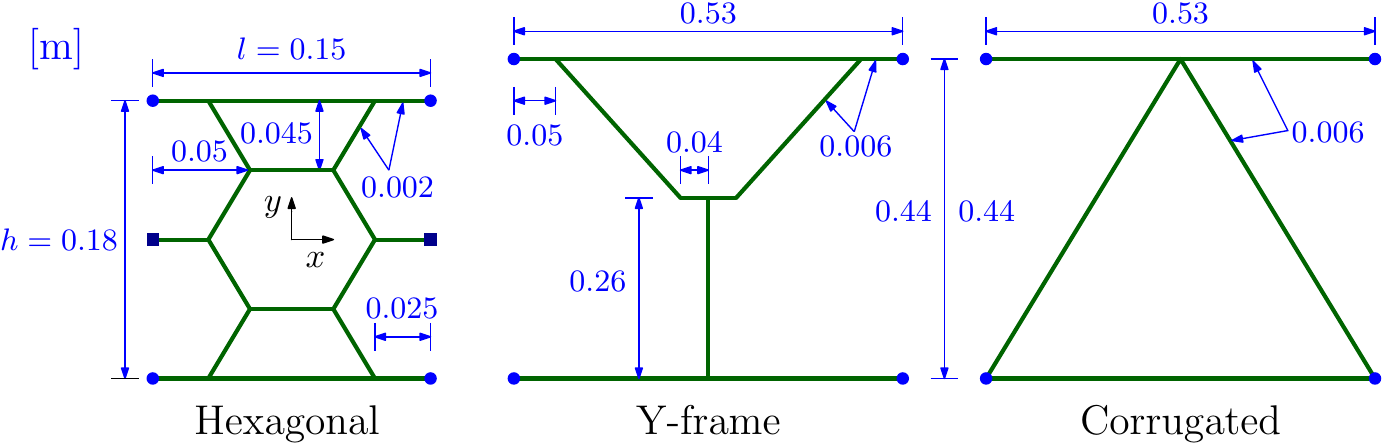}
\caption{Unit cells of hexagonal, Y-frame and corrugated lattice core sandwich beams. The latter two are modeled according to \cite{luc2015}}
\end{figure}

For the other cores displayed in Fig.~2, Young's modulus and Poisson ratio are $E=210$ GPa and $\nu=0.3$, respectively. Other relevant parameters are given in Fig.~2. The hexagonal core includes two mid-nodes in addition to the four corner nodes that need to be taken into account essentially to ensure connectivity between neighboring unit cells on the micropolar continuum level. Static condensation is applied at the inner nodes of the unit cell. It is difficult to obtain a meaningful symbolic form for the hexagonal constitutive matrix, in numerical form we have
\begin{equation}
\mathbf{C}_{\textrm{hex}}=
\begin{bmatrix}
 4.26438\cdot10^7 & 0 & 0 & 0 & 0 \\
 0 & 340740 & 0 & 0 & 27.1761  \\
 0 & 0 & 94735.4 & 8319.08 & 0 \\
 0 & 0 & 8319.08 & 3315.25 & 0  \\
 0 & 27.1761 & 0 & 0 & 44.3467 \\
\end{bmatrix}
\end{equation}
The constitutive matrix (25) is of the same form as that of the web-core with the exception that for the hexagonal core $C_{33}\neq C_{44}$. For the Y-frame and corrugated cores we obtain
\begin{equation}
\mathbf{C}_{\textrm{Y}}=
\begin{bmatrix}
 1.26053\cdot10^8 & 11696.2 & 0 & 0 & 5292.50 \\
 11696.2 & 6.10097\cdot10^6 & 0 & 0 & 1164.35  \\
 0 & 0 & 42094.9 & 9541.42 & 0 \\
 0 & 0 & 9541.42 & 5302.30 & 0  \\
 5292.50 & 1164.35 & 0 & 0 & 1012.92 \\
\end{bmatrix}
\end{equation}
and
\begin{equation}
\mathbf{C}_{\textrm{corr}}=
\begin{bmatrix}
 1.26018\cdot10^8 & -3902.45 & 0 & 0 & -3902.45 \\
 -3902.45 & 6.09926\cdot10^6 & 0 & 0 & 858.539  \\
 0 & 0 & 2.09792\cdot10^7 & 3734.96 & 0 \\
 0 & 0 & 3734.96 & 5078.42 & 0  \\
 -3902.45 & 858.539 & 0 & 0 & 1334.05 \\
\end{bmatrix},
\end{equation}
respectively. The axial and classical sandwich bending stiffnesses in Eqs.~(25)--(27) are practically given by $C_{11}\approx2EA_f$ and $C_{22}\approx EA_f h^2/2$, respectively. We see that due to the lack of symmetry about the $x$-axis, the coupling terms $C_{12}$ and $C_{15}$ appear in the constitutive matrices of the Y-frame and corrugated cores. In addition, the symmetric shear stiffness $C_{33}$ of the corrugated core is very high in comparison to that of the other cores because the corrugated lattice core has a stretch-dominated shear-carrying mechanism while the others cores are bending-dominated. In other words, when bent, the constituents of the corrugated lattice core act as axial rods without significant bending so that the core is very stiff. The differences between bending- and stretch-dominated cores in geometrically nonlinear bending problems will be studied further by numerical examples in Section 4. Finally, it is easy verify that all eigenvalues of each constitutive matrix above are positive which means that the matrices are positive definite. It follows that each lattice core material is stable in the conventional sense (i.e., strain energy is positive for nonzero strains).
\subsection{Geometrically nonlinear beam equations}
The principle of virtual displacements can be stated for a geometrically nonlinear micropolar Timoshenko beam as
\begin{equation}
\delta W=\delta W_I-\delta W_E=0,
\end{equation}
The virtual strain energy stored in a typical beam element, $\Omega=(x_{a},x_{b})$, and consistent with the strains used in the previous section, is
\begin{equation}
\begin{aligned}
\delta W_I&=\int_V(\sigma_x\delta{\hat{\varepsilon}}_x+m_{xz}\delta\kappa_{xz}+\tau_{xy}\delta\varepsilon_{xy}+\tau_{yx}\delta\varepsilon_{yx})dV \\
&=\int_{x_a}^{x_b}\big[N_x(\delta u_x'+u_y'\delta u_y')+M_x\delta\phi'+P_{xz}\delta\psi'\\
&\qquad\quad+Q_{xy}(\delta u_y'-\delta\psi)+Q_{yx}(\delta\phi+\delta\psi)\big]dx,
\end{aligned}
\end{equation}
where $\hat{\varepsilon}_x=\hat{\varepsilon}_x^0+y\kappa_x$ [cf. Eqs.~(2) and (19)]. The external virtual work is
\begin{equation}
\delta W_E=\int_{x_a}^{x_b}q\delta u_y\,dx+\sum_{i=1}^{8}Q_i\delta U_i,
\end{equation}
where $q(x)$ is the distributed transverse load, $Q_i$ are the generalized external forces at the beam ends and $U_i$ are the associated generalized displacements defined as
\begin{equation}
\begin{aligned}
        U_1&=u_x(x_a), & \qquad U_5&=u_x(x_b), \\
        U_2&=u_y(x_a), & \qquad U_6&=u_y(x_b), \\
        U_3&=\phi(x_a), & \qquad U_7&=\phi(x_b) \\
        U_4&=-\psi(x_a), & \qquad U_8&=-\psi(x_b).
\end{aligned}
\end{equation}
Equation (28) conforms to the micropolar beam finite element presented in Fig.~3 and provides the weak form for the finite element formulation in Section 3. On the other hand, Eq.~(28) yields the Euler-Lagrange equations
\begin{align}
N_x'=0, \quad V_{xy}'=-q, \quad Q_{yx}-M_x'=0, \quad
P_{xz}'+Q_{xy}-Q_{yx}=0,
\end{align}
where the effective transverse shear force is
\begin{equation}
V_{xy}=Q_{xy}+N_x u_y'.
\end{equation}
Finally, when the corresponding generalized displacements are not defined, the natural (or force) boundary conditions become
\begin{equation}
\begin{aligned}
        Q_1&=-N_x(x_a), & \qquad Q_5&=N_x(x_b), \\
        Q_2&=-V_{xy}(x_a), & \qquad Q_6&=V_{xy}(x_b), \\
        Q_3&=-M_x(x_a), & \qquad Q_7&=M_x(x_b), \\
        Q_4&=P_{xz}(x_a), & \qquad Q_8&=-P_{xz}(x_b).
\end{aligned}
\end{equation}
The equilibrium equations (32) may be presented in terms of displacements by using the constitutive relations (21). The governing equations could also have been derived by employing the principle of minimum total potential energy and the strain energy density $\hat{W}_0^l$ given in Eq.~(20).
\begin{figure}
\centering
\includegraphics[scale=1.3]{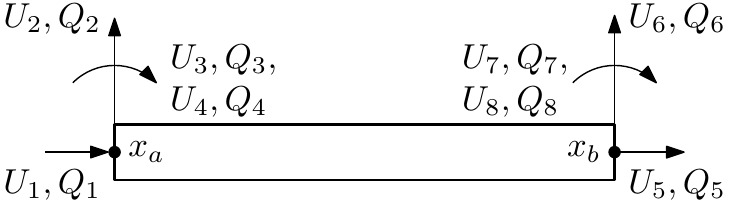}
\caption{Set-up according to which the micropolar Timoshenko beam finite element will be developed in Section 3.}
\end{figure}
\section{Geometrically nonlinear micropolar Timoshenko beam element}
\subsection{Finite element formulation}
The finite element model is developed using the statement of the principle of virtual displacements (28) over a typical element $\Omega^{e} = (x_a,x_b) $, which takes the explicit form
\begin{align}
0&=\int_{x_a}^{x_b}\left\{N_x(\delta u_x'+u_y'\delta u_y')+M_x\delta\phi'+P_{xz}\delta\psi'
+Q_{xy}(\delta u_y'-\delta\psi)+Q_{yx}(\delta\phi+\delta\psi)\right\}dx \nonumber \\
&-\int_{x_a}^{x_b}q\delta u_y\,dx-\sum_{i=1}^{8}Q_i\delta U_i
\end{align}
After using the constitutive equations in (21), the following four statements for a typical finite element $\Omega^{e}=(x_a,x_b)$ which are equivalent to Eq.~(35) can be written
\begin{align}
    0&=\int_{x_a}^{x_b}\left\{C_{11}\left(u_x'+\frac{1}{2}\left(u_y'\right)^2\right)+C_{12}\phi'+C_{15}\psi'\right\}\delta u_x'\,dx-Q_1\delta U_1-Q_5\delta U_5 \\
    0&=\int_{x_a}^{x_b}\Biggl\{\left(\left[C_{11}\left(u_x'+\frac{1}{2}\left(u_y'\right)^2\right)+C_{12}\phi'+C_{15}\psi'\right]u_y'\right)\delta u_y' \nonumber \\
    &\qquad\quad+\left(\left(C_{33}+C_{34}\right)\left(u_y'+\phi\right)+\left(C_{34}+C_{44}\right)\left(u_y'-\phi-2\psi\right)\right)\delta u_y'\Biggr\}dx \nonumber \\
    &\qquad\quad-\int_{x_a}^{x_b}q\delta u_ydx-Q_2\delta U_2-Q_6\delta U_6 \\
    0&=\int_{x_a}^{x_b}\Biggl\{\left(C_{12}\left(u_x'+\frac{1}{2}\left(u_y'\right)^2\right)+C_{22} \phi'+C_{25} \psi'\right)\delta\phi' \nonumber \\
    &\qquad\quad+\left(\left(C_{33}-C_{34}\right)(u_y'+\phi)+(C_{34}-C_{44})(u_y'-\phi-2\psi)\right)\delta \phi\Biggr\}dx \nonumber \\
    &\qquad\quad-Q_3\delta U_3-Q_7\delta U_7
\end{align}
\begin{align}
    0&=\int_{x_a}^{x_b}\Biggl\{\left(C_{15}\left(u_x'+\frac{1}{2}\left(u_y'\right)^2\right)+C_{25}\phi'+C_{55}\psi'\right)\delta \psi' \nonumber \\
    &\qquad\quad-2\left(C_{34}(u_y'+\phi)+C_{44}(u_y'-\phi-2\psi\right)\delta \psi\Biggr\}dx \nonumber \\
    &\qquad\quad-Q_4\delta U_4-Q_8\delta U_8
\end{align}
The primary variables are $u_x,u_y,\phi$ and $\psi$. These variables are approximated using Lagrange interpolation functions $L_j^{(J)}$, $(J=1,2,3,4)$ (see, for example, \cite{reddy2019}) so that
\begin{equation}
\begin{aligned}
    u_x& \approx\sum_{j=1}^{2}u_jL_j^{(1)}(x) , \quad
    u_y\approx\sum_{j=1}^{2}v_jL_j^{(2)}(x) \\
    \phi&\approx\sum_{j=1}^{2}\phi_jL_j^{(3)}(x), \quad
  \  \psi\approx\sum_{j=1}^{2}\psi_jL_j^{(4)}(x) \\
\end{aligned}
\end{equation}
By substituting Eq.~(40) for $u_x,u_y,\phi$ and $\psi$ and putting $\delta u_x = L_j^{(1)} ,\delta u_y = L_j^{(2)} ,\delta \phi = L_j^{(3)} $ and $\delta \psi = L_j^{(4)}$ into the weak-form statements in Eqs.~(36)--(39), the finite element equations for a typical beam element can be expressed as
\begin{equation}
\begin{bmatrix}
    \mathbf{K^{11}}&\mathbf{K^{12}}&\mathbf{K^{13}}&\mathbf{K^{14}} \\
    \mathbf{K^{21}}&\mathbf{K^{22}}&\mathbf{K^{23}}&\mathbf{K^{24}} \\
    \mathbf{K^{31}}&\mathbf{K^{32}}&\mathbf{K^{33}}&\mathbf{K^{34}} \\
    \mathbf{K^{41}}&\mathbf{K^{42}}&\mathbf{K^{43}}&\mathbf{K^{44}} \\
\end{bmatrix}^{(e)}
\begin{Bmatrix}
    \mathbf{u_x} \\
    \mathbf{u_y} \\
    \boldsymbol{\phi} \\
    \boldsymbol{\psi} \\
\end{Bmatrix}^{(e)}
=
\begin{Bmatrix}
    \mathbf{F^1} \\
    \mathbf{F^2} \\
    \mathbf{F^3} \\
    \mathbf{F^4} \\
\end{Bmatrix}^{(e)}
\end{equation}
The stiffness coefficients $K_{ij}^{\alpha\beta}$ and force coefficients $F_{i}^{\alpha}$ ($\alpha,\beta =1,2,3,4$ and $i,j=1,2$) are defined as
\begin{equation}
\begin{aligned}
    K_{ij}^{11}&=\int_{x_a}^{x_b}\left\{C_{11}\frac{dL_i^{(1)}}{dx}\frac{dL_j^{(1)}}{dx}\right\}dx \nonumber \\
    K_{ij}^{12}&=\frac{1}{2}\int_{x_a}^{x_b}\left\{C_{11}\left(\frac{du_y}{dx}\right)\frac{dL_i^{(1)}}{dx}\frac{dL_j^{(2)}}{dx}\right\}dx \nonumber \\
    K_{ij}^{13}&=\int_{x_a}^{x_b}\left\{C_{12}\frac{dL_i^{(1)}}{dx}\frac{dL_j^{(3)}}{dx}\right\}dx \nonumber \\
    K_{ij}^{14}&=\int_{x_a}^{x_b}\left\{C_{15}\frac{dL_i^{(1)}}{dx}\frac{dL_j^{(4)}}{dx}\right\}dx \nonumber \\
    K_{ij}^{21}&=\int_{x_a}^{x_b}\left\{C_{11}\left(\frac{du_y}{dx}\right)\frac{dL_i^{(2)}}{dx}\frac{dL_j^{(1)}}{dx}\right\}dx \nonumber
\end{aligned}
\end{equation}
\begin{equation}
\begin{aligned}
    K_{ij}^{22}&=\frac{1}{2}\int_{x_a}^{x_b}\left\{C_{11}\left(\frac{du_y}{dx}\right)^2\frac{dL_i^{(2)}}{dx}\frac{dL_j^{(2)}}{dx}\right\}dx + \int_{x_a}^{x_b}\left\{\left(C_{33}+2C_{34}+C_{44}\right)\frac{dL_i^{(2)}}{dx}\frac{dL_j^{(2)}}{dx}\right\}dx \\
    K_{ij}^{23}&=\int_{x_a}^{x_b}\left\{C_{12}\left(\frac{du_y}{dx}\right)\frac{dL_i^{(2)}}{dx}\frac{dL_j^{(3)}}{dx}\right\}dx + \int_{x_a}^{x_b}\left\{\left(C_{33}-C_{44}\right)\frac{dL_i^{(2)}}{dx}L_j^{(3)}\right\}dx \\
    K_{ij}^{24}&=\int_{x_a}^{x_b}\left\{C_{15}\left(\frac{du_y}{dx}\right)\frac{dL_i^{(2)}}{dx}\frac{dL_j^{(4)}}{dx}\right\}dx - 2\int_{x_a}^{x_b}\left\{(C_{34}+C_{44})\frac{dL_i^{(2)}}{dx}L_j^{(4)}\right\}dx\\
    K_{ij}^{31}&=\int_{x_a}^{x_b}\left\{C_{12}\frac{dL_i^{(3)}}{dx}\frac{dL_j^{(1)}}{dx}\right\}dx \\
    K_{ij}^{32}&=\frac{1}{2}\int_{x_a}^{x_b}\left\{C_{12}\left(\frac{du_y}{dx}\right)\frac{dL_i^{(3)}}{dx}\frac{dL_j^{(2)}}{dx}\right\}dx + \int_{x_a}^{x_b}\left\{\left(C_{33}-C_{44}\right)L_i^{(3)}\frac{dL_j^{(2)}}{dx}\right\}dx \\
    K_{ij}^{33}&=\int_{x_a}^{x_b}\left\{C_{22}\frac{dL_i^{(3)}}{dx}\frac{dL_j^{(3)}}{dx}\right\}dx + \int_{x_a}^{x_b}\left\{(C_{33}+C_{44})L_i^{(3)}L_j^{(3)}\right\}dx \\
    K_{ij}^{34}&=\int_{x_a}^{x_b}\left\{C_{25}\frac{dL_i^{(3)}}{dx}\frac{dL_j^{(4)}}{dx}\right\}dx + 2\int_{x_a}^{x_b}\left\{(C_{44}-C_{34})L_i^{(3)}L_j^{(4)}\right\}dx \\
    K_{ij}^{41}&=\int_{x_a}^{x_b}\left\{C_{15}\frac{dL_i^{(4)}}{dx}\frac{dL_j^{(1)}}{dx}\right\}dx \\
    K_{ij}^{42}&=\frac{1}{2}\int_{x_a}^{x_b}\left\{C_{15}\left(\frac{du_y}{dx}\right)\frac{dL_i^{(4)}}{dx}\frac{dL_j^{(2)}}{dx}\right\}dx - 2\int_{x_a}^{x_b}\left\{\left(C_{34}+C_{44}\right)L_i^{(4)}\frac{dL_j^{(2)}}{dx}\right\}dx \\
    K_{ij}^{43}&=\int_{x_a}^{x_b}\left\{C_{25}\frac{dL_i^{(4)}}{dx}\frac{dL_j^{(3)}}{dx}\right\}dx - 2\int_{x_a}^{x_b}\left\{(C_{34}-C_{44})L_i^{(4)}L_j^{(3)}\right\}dx \\
    K_{ij}^{44}&=\int_{x_a}^{x_b}\left\{C_{55}\frac{dL_i^{(4)}}{dx}\frac{dL_j^{(4)}}{dx}\right\}dx + 4\int_{x_a}^{x_b}\left\{C_{44}L_i^{(4)}L_j^{(4)}\right\}dx \\
\end{aligned}
\end{equation}
\begin{equation}
\begin{aligned}
    F_{i}^{1}&= Q_1L_{i}^{(1)}(x_a) + Q_5L_{i}^{(1)}(x_b) \\
    F_{i}^{2}&= \int_{x_a}^{x_b}q(x)L_{i}^{(2)}dx + Q_2L_{i}^{(1)}(x_a) + Q_6L_{i}^{(1)}(x_b) \\
    F_{i}^{3}&= Q_3L_{i}^{(1)}(x_a) + Q_7L_{i}^{(1)}(x_b) \\
    F_{i}^{4}&= Q_4L_{i}^{(1)}(x_a) + Q_8L_{i}^{(1)}(x_b)
\end{aligned}
\end{equation}
\subsection{Solution of nonlinear equations}
The nonlinear finite element equations (41) are solved iteratively using the Newton's iteration procedure (see \cite{reddy2015}). The linearized element equation at the beginning of the $r^{th}$ iteration will take the form:
\begin{align}
    \mathbf{T}^{(e)}(\mathbf{U}^{(e)(r-1)})\Delta\mathbf{U}^{(e)(r)}& = -\mathbf{R}^{(e)}(\mathbf{U}^{(e)(r-1)})
\end{align}
Where $\mathbf{U}^{(e)(r-1)}$ is the generalized nodal displacement vector of element $e$ at the end of $(r-1)^{th}$ iteration and $\Delta\mathbf{U}^{(e)(r)}$ is the incremental displacement vector of element $e$ at the $r^{th}$ iteration defined as
\begin{align}
    \mathbf{U}^{(e)(r)}=\mathbf{U}^{(e)(r-1)}+\Delta\mathbf{U}^{(e)(r)}
\end{align}
The residual vector $\mathbf{R}^{(e)}(\mathbf{U}^{(e)(r-1)})$, computed after the end of $(r-1)^{th}$ iteration, is defined as
\begin{align}
    \mathbf{R}^{(e)}(\mathbf{U}^{(e)(r-1)})=\mathbf{K}^{(e)}\mathbf{U}^{(e)(r-1)}-\mathbf{F}^{(e)}
\end{align}
Once the residual vector is computed using Eq.~(46), the tangent stiffness matrix $\mathbf{T}^{(e)}$ can be calculated using the following definition
\begin{align}
    \mathbf{T}^{(e)}\equiv\frac{\partial{\mathbf{R}^{(e)}}}{\partial{\mathbf{U}^{(e)}}}\quad \textnormal{or}\quad T_{ij}^{(e)}=\frac{\partial{R}_i^{(e)}}{\partial{U}_j^{(e)}}
\end{align}
For a typical element, writing Eq.~(44) in a fashion similar to Eq.~(41), we get
\begin{equation}
\begin{bmatrix}
    \mathbf{T^{11}}&\mathbf{T^{12}}&\mathbf{T^{13}}&\mathbf{T^{14}} \\
    \mathbf{T^{21}}&\mathbf{T^{22}}&\mathbf{T^{23}}&\mathbf{T^{24}} \\
    \mathbf{T^{31}}&\mathbf{T^{32}}&\mathbf{T^{33}}&\mathbf{T^{34}} \\
    \mathbf{T^{41}}&\mathbf{T^{42}}&\mathbf{T^{43}}&\mathbf{T^{44}} \\
\end{bmatrix}^{(e)}
\begin{Bmatrix}
    \Delta\mathbf{U^1} \\
    \Delta\mathbf{U^2} \\
    \Delta\mathbf{U^3} \\
    \Delta\mathbf{U^4} \\
\end{Bmatrix}^{(e)}
=-
\begin{Bmatrix}
    \mathbf{R^1} \\
    \mathbf{R^2} \\
    \mathbf{R^3} \\
    \mathbf{R^4} \\
\end{Bmatrix}^{(e)}
\end{equation}
where the notation $\mathbf{U}^1= \mathbf{u_x}$, $\mathbf{U}^2= \mathbf{u_y}$, $\mathbf{U}^3=\boldsymbol{\phi}$ and $\mathbf{U}^4=\boldsymbol{\psi}$ is used. Then the coefficients of the tangent stiffness matrix, $T_{ij}^{\alpha\beta}$ ($\alpha,\beta=1,2,3,4$ and $i,j=1,2$), in the above equation can be computed as
\begin{align}
   T_{ij}^{\alpha\beta}=K_{ij}^{\alpha\beta}+\sum_{\gamma=1}^{4}\sum_{p=1}^{2}\frac{\partial(K_{ip}^{\alpha\gamma})}{\partial{U_{j}^{\beta}}}U_{p}^{\gamma}
\end{align}
In explicit terms, the coefficients read
\begin{equation}
\begin{aligned}
    T_{ij}^{11}&=K_{ij}^{11},\quad T_{ij}^{12}=K_{ij}^{12}+\frac{1}{2}\int_{x_a}^{x_b}\left\{C_{11}\frac{du_y}{dx}\frac{dL_i^{(1)}}{dx}\frac{dL_j^{(2)}}{dx}\right\}dx,\quad
    T_{ij}^{13}=K_{ij}^{13},\quad
    T_{ij}^{14}=K_{ij}^{14},\quad \\
\end{aligned}
\end{equation}
\begin{equation}
\begin{aligned}
    T_{ij}^{21}&=K_{ij}^{21},\quad
    T_{ij}^{23}=K_{ij}^{23},\quad
    T_{ij}^{24}=K_{ij}^{24},\quad \\
    T_{ij}^{31}&=K_{ij}^{31},\quad T_{ij}^{32}=K_{ij}^{32}+\frac{1}{2}\int_{x_a}^{x_b}\left\{C_{12}\frac{du_y}{dx}\frac{dL_i^{(3)}}{dx}\frac{dL_j^{(2)}}{dx}\right\}dx,\quad
    T_{ij}^{33}=K_{ij}^{33},\quad
    T_{ij}^{34}=K_{ij}^{34},\quad \\
    T_{ij}^{41}&=K_{ij}^{41},\quad T_{ij}^{42}=K_{ij}^{42}+\frac{1}{2}\int_{x_a}^{x_b}\left\{C_{15}\frac{du_y}{dx}\frac{dL_i^{(4)}}{dx}\frac{dL_j^{(2)}}{dx}\right\}dx,\quad
    T_{ij}^{43}=K_{ij}^{43},\quad
    T_{ij}^{44}=K_{ij}^{44},\quad \\
    T_{ij}^{22}&=K_{ij}^{22}+\int_{x_a}^{x_b}\left\{C_{11}\frac{du_x}{dx}\frac{dL_i^{(2)}}{dx}\frac{dL_j^{(2)}}{dx}\right\}dx+\int_{x_a}^{x_b}\left\{C_{11}{\left(\frac{du_y}{dx}\right)}^2\frac{dL_i^{(2)}}{dx}\frac{dL_j^{(2)}}{dx}\right\}dx\\
    &\qquad+\int_{x_a}^{x_b}\left\{C_{12}\frac{d\phi}{dx}\frac{dL_i^{(2)}}{dx}\frac{dL_j^{(2)}}{dx}\right\}dx
    +\int_{x_a}^{x_b}\left\{C_{15}\frac{d\psi}{dx}\frac{dL_i^{(2)}}{dx}\frac{dL_j^{(2)}}{dx}\right\}dx\\
\end{aligned}
\end{equation}
The element equations computed from Eq.~(48) are assembled and solved (after the imposition of the boundary conditions) to obtain the global incremental displacement vector $\Delta\mathbf{U}$ at the $r^{th}$ iteration. The normalized difference between solution vectors from two consecutive iterations, measured with Euclidean norm, is computed at the end of each iteration. If the value computed is less than a preselected tolerance $\varepsilon$ further iterations are terminated
\begin{align}
    \sqrt{\frac{\Delta\mathbf{U}\cdot\Delta\mathbf{U}}{\mathbf{U}^{(r)}\cdot\mathbf{U}^{(r)}}}=\sqrt{\frac{\sum_{I=1}^{N}{|U_I^{(r)}-U_I^{(r-1)}|}^2}{\sum_{I=1}^{N}{|U_I^{(r)}|}^2}}\leqslant\varepsilon
\end{align}
Further, acceleration of convergence for some type of nonlinearities may be achieved by using weighted-average of solutions from the last two iterations rather than the solution from the last iteration to evaluate the coefficient matrix:
\begin{align}
    \mathbf{U}^{(r)}={\mathbf{K}(\mathbf{\overline{U}})}^{-1}\mathbf{F}(\mathbf{\overline{U}}),\quad \mathbf{\overline{U}}\equiv\beta\mathbf{U}^{(r-2)}+(1-\beta)\mathbf{U}^{(r-1)},\quad 0\leqslant\beta\leqslant1
\end{align}
where $\beta$ is known as the acceleration parameter. The value of $\beta$ depends on the nature of nonlinearity and the type of problem considered.
Once the convergence is obtained the final solution is computed using
\begin{align}
    \mathbf{U}^{(r)}=\mathbf{U}^{(r-1)}+\Delta\mathbf{U}^{(r)}
\end{align}
\subsection{Shear and Membrane locking}
The finite element model used in this paper uses linear interpolation on both $u_y$ and $\phi$. However, in the thin beam limit, when linear interpolation is used for $u_y$, the cross-sectional rotation $\phi$ should approach $-\left(du_y/dx\right)$, which is necessarily constant. But since $\phi$ is also interpolated as linear, it can never be constant. This inconsistency causes what is known as shear locking (see \cite{reddy2015}). To avoid this inconsistency, we may use equal interpolation on both $u_y$ and $\phi$ but treat $\phi$ as constant while evaluating the symmetric $\gamma_s$ and anti-symmetric $\gamma_a$ shear strains. This amounts to using reduced Gauss quadrature rule in evaluating the integrals containing constants $C_{33}$, $C_{34}$ and $C_{44}$ while computing the element coefficient matrices of Eq.~(41) and Eq.~(48).

When von K\'arm\'an nonlinearity is included, there is coupling between $u_x$ and $u_y$ which causes the beam to undergo axial displacement even when there are no axial forces. But in the case of hinged-hinged beam, there are no constraints on $u_x$ at the boundaries, thus causing the beam to roll over freely without axial strain, i.e,
\begin{align*}
    \hat{\varepsilon}_{x}^0=\frac{du_x}{dx}+\frac{1}{2}\left(\frac{du_y}{dx}\right)^2=0
\end{align*}
In order to satisfy this we need
\begin{align*}
    -\frac{du_x}{dx}\sim\left(\frac{du_y}{dx}\right)^2
\end{align*}
In essence, we need to have the same degree of polynomial variation on both $\left(du_x/dx\right)$ and $\left(du_y/dx\right)^2$. But when equal interpolation of degree greater than one is used for both $u_x$ and $u_y$ this criteria cannot be satisfied and leads to what is known as membrane locking (see \cite{reddy2015}). To overcome this we have to treat $\left(du_y/dx\right)^2$ as same order as $\left(du_x/dx\right)$. This is achieved using reduced integration while evaluating all the nonlinear terms of the element coefficients matrices of Eqs.~(41) and (48).
\section{Numerical bending examples}
\subsection{General Setup}
The developed micropolar beam finite element model is used for bending analysis of lattice core sandwich beams. The four structural cores considered in Section 2.3 are used in the calculations. Both geometrically linear and nonlinear cases are analyzed using the 1-D beam model. 2-D reference solutions are computed using Euler--Bernoulli FE beam frames modeled by Abaqus; the pins in simply-supported cases are at the central axis of the 2-D frame so that the model corresponds to 1-D cases.
\subsection{Bending of a web-core beam}
A beam consisting of 24 web-core unit cells is considered first. The length of each web-core unit cell is $l=0.12$ m resulting in a total beam length of $L=2.88$ m. The beam is analyzed for two different boundary conditions, namely, a fixed-fixed case and a three-point-bending setup. For the fixed-fixed case the boundaries are subjected to the following conditions:
\begin{equation}
\begin{aligned}
    x=0: u_x=0 , u_y = 0, \phi =0, \psi = 0 \\
    x=L: u_x=0 , u_y = 0, \phi =0, \psi = 0
\end{aligned}
\end{equation}
\begin{figure}
    \centering
    \includegraphics[scale=1.2,trim=0 30 0 40,clip]{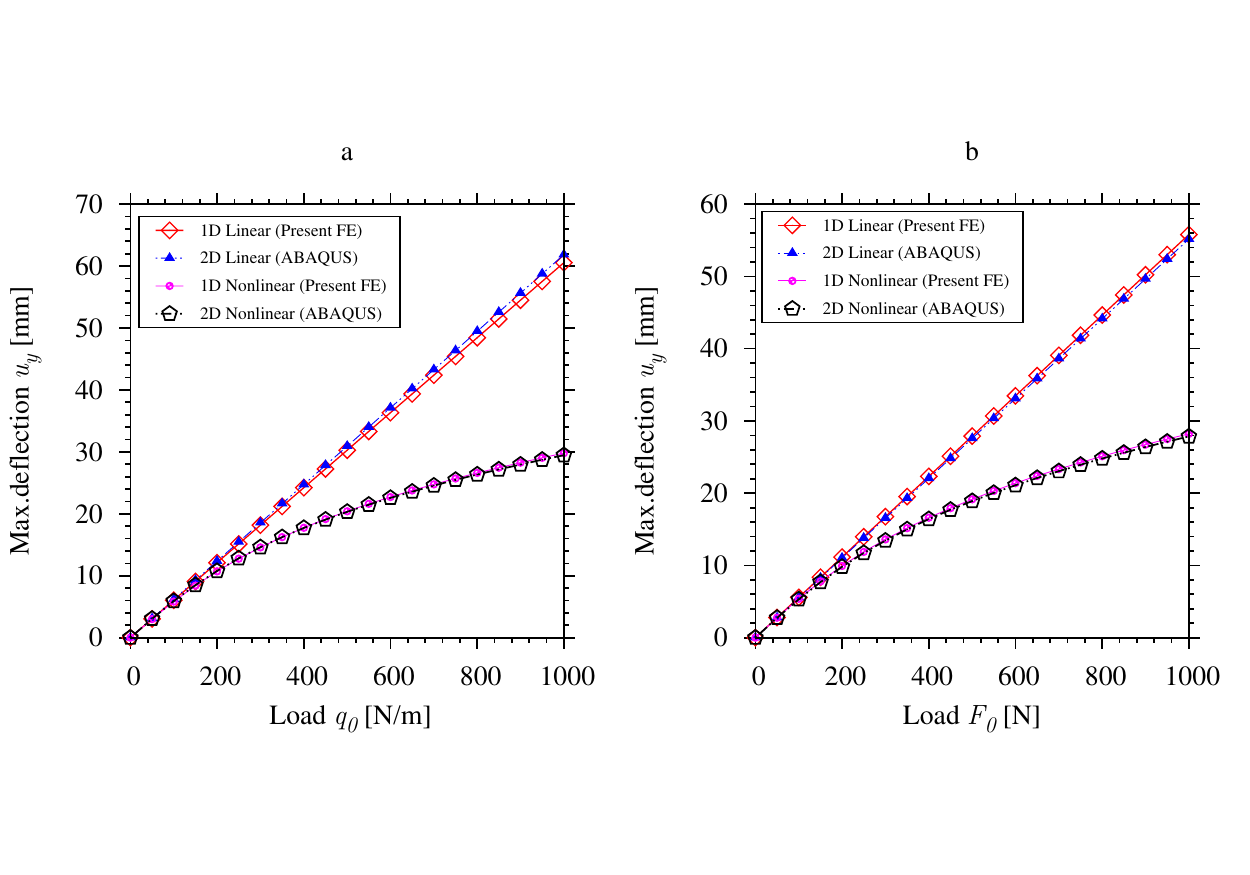}
    \caption{(a) Maximum transverse deflection of a fixed-fixed web-core beam subjected to a uniformly distributed load. (b) Maximum transverse deflection of a web-core beam under three-point-bending.}
\end{figure}
A uniformly distributed load $q_0$ is exerted on the beam. The load is applied in increments of $\Delta q_0=50$ N/m until a maximum load of $1000$ N/m is reached. The maximum deflection, which occurs at the center of the beam, is recorded against the corresponding applied load. The results for both linear and nonlinear cases are plotted in Fig.~4a. The nonlinear deflections are smaller than the linear deflections at large loads because, as the load increases, the internal forces resisting the deformation increase in a nonlinear fashion.

For the three-point-bending case the boundaries are subjected to the following conditions:
\begin{equation}
\begin{aligned}
    x=0: u_x=0 , u_y = 0, M_x =0, P_{xz} = 0 \\
    x=L: u_x=0 , u_y = 0, M_x =0, P_{xz} = 0
\end{aligned}
\end{equation}
Here, instead of a uniformly distributed load, a point load $F_0$ is applied at the center of the beam. The point load is applied in increments of $\Delta F_0=50$ N until a maximum load of $1000$ N is reached. The maximum deflection, which occurs at the center of the beam, is recorded against the corresponding applied load. The results from the finite element model developed here for the 1-D  equivalent single layer beam are compared with the 2-D FE results (see Fig.~4b). Note that ABAQUS uses a more complete Green-Lagrange strain tensor for the geometrically nonlinear beam element, whereas in the present finite element model developed in this paper the nonlinearity is included in the form of von K\'arm\'an strains.
\begin{figure}
    \centering
    \includegraphics[scale=1.2,trim=0 30 0 40,clip]{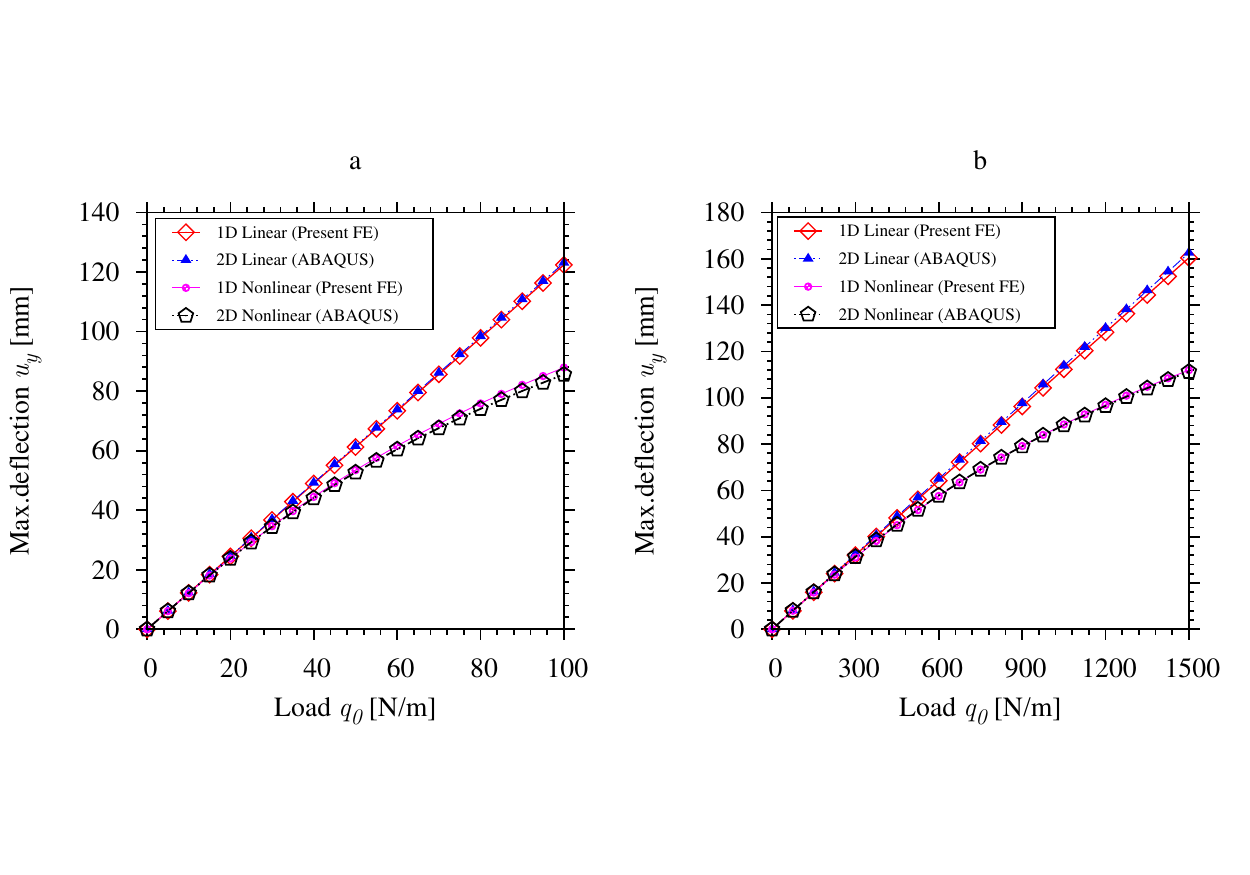}
    \caption{Maximum deflection of a (a) Y-frame and (b) hexagonal core sandwich beams subjected to a uniformly distributed load under fixed-fixed boundary conditions.}
\end{figure}
\subsection{Fixed-fixed hexagonal and Y-frame core beams}
Here we consider two beams, one made of 48 hexagonal core unit cells and the other made of 30 Y-frame unit cells (see Fig.~2). Thus, the total length of the hexagonal core beam is $L=7.2$ m and the length of Y-frame core beam is $L=15.9$ m. Both the beams are subjected to a uniformly distributed load $q_0$. Fixed-fixed boundary conditions (55) are applied at the beam ends. For the hexagonal core beam the load is applied in increments of $\Delta q_0=75$ N/m until a maximum load of $1500$ N/m is reached, while for the Y-frame core beam the load is applied in increments of $\Delta q_0=5$ N/m until a maximum load of $100$ N/m is reached. The maximum transverse deflections, which occur at the beam centers, are recorded and plotted against the corresponding applied load in Figs.~5a and 5b.
\begin{figure}
    \centering
    \includegraphics[scale=1.2,trim=0 30 0 40,clip]{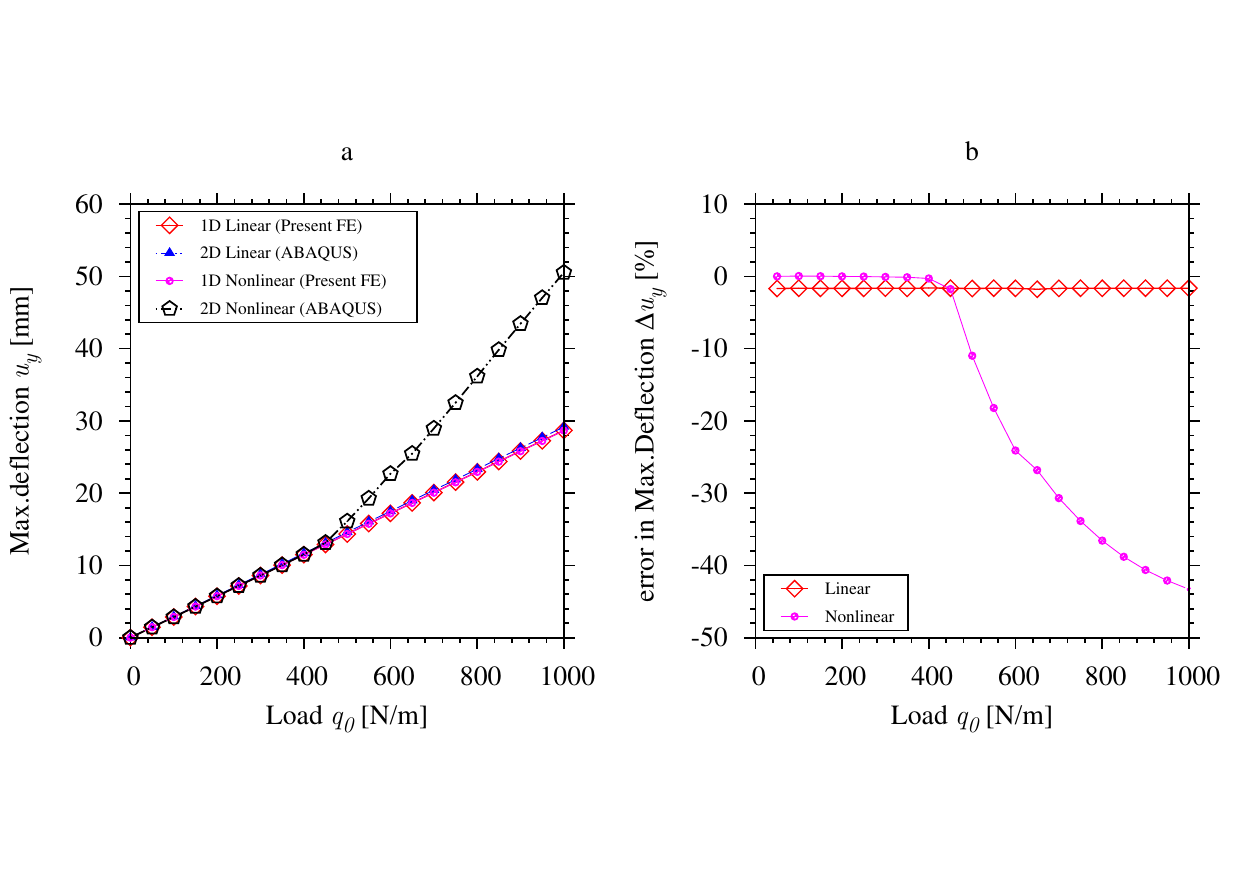}
    \caption{(a) Maximum deflection of a corrugated core sandwich beam subjected a uniformly distributed load under fixed-fixed boundary conditions. Local buckling occurs in the 2-D reference model near $q=500$ N/m which cannot be accounted for by the micropolar 1-D model. (b) Percentage error of 1D beam model developed, in terms of maximum vertical deflection relative to 2D-beam frame solution (face sheet deflection) calculated using ABAQUS.  }
\end{figure}
\subsection{Fixed-fixed corrugated core beam}
A beam consisting of 30 corrugated core unit cells is considered. Since the length of each corrugated unit cell is $l=0.53$ m, the total length of the beam is $L=15.9$ m (see Fig.~2). The beam is subjected to fixed-fixed boundary conditions (55). A uniformly distributed load $q_0$ is applied on the beam. The load is applied in increments of $\Delta q_0=50$ N/m until a maximum load of $1000$ N/m is reached. The maximum vertical deflection of the beam is plotted against the corresponding applied load in Fig.~6a. The error in the maximum vertical deflection is calculated using,
\begin{align}
    \Delta u_y = 100\times\left(\frac{u_y^{1-D\text{ micropolar}}-u_y^{2-D\text{ beam frame}}}{u_y^{2-D\text{ beam frame}}}\right)
\end{align}
and is plotted against the applied load for both the linear and nonlinear cases in Fig.~6b.

Unlike the other structural cores, we see that the nonlinear deflections of the corrugated core beam, calculated using the finite element model developed for the 1-D equivalent single layer beam, are not in good agreement with the 2-D beam frame results all the way. This is due to the local buckling of the stretch-dominated corrugated core that occurs in the 2-D model as displayed in Fig.~7. The presented 1-D equivalent single layer model cannot account for this local buckling.

We also note that even though the lengths and heights of both corrugated core and Y-frame core beams are equal, the corrugated core beam is much stiffer than the Y-frame core beam. The maximum nonlinear deflection for the Y-frame core beam subjected to fixed-fixed boundary conditions is $88$ mm at a uniformly distributed load of $100$ N/m (see Fig.~5a), while for the corrugated core beam the maximum deflection for a uniformly distributed load of $100$ N/m is only $2.9$ mm (see Fig.~6a). The high stiffness of the corrugated core beam is attributed to its stretch-dominated behavior unlike Y-frame core which is bending-dominated. The corrugated core has a very high shear stiffness because of the fact that the elements (the Euler-Bernoulli beam elements within the core structure) of the corrugated core act essentially like rods and do not bend much, where as this is not the case in the Y-frame core. Although the elements of the part which resemble the corrugated core (the upper `V' part of `Y') in the Y-frame core do not exhibit lot of bending, the remaining part, consisting of lower element, undergoes significant bending.
\begin{figure}
    \centering
    \includegraphics[scale=0.4]{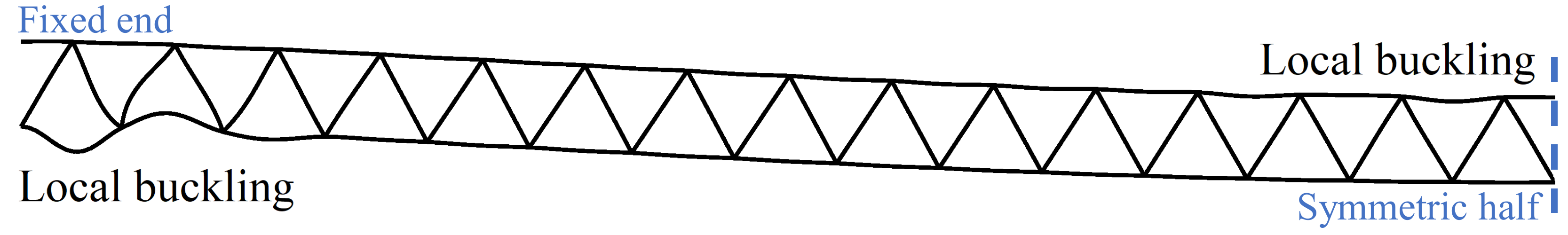}
    \caption{Local buckling of a corrugated core sandwich beam modeled as a 2-D FE beam frame (Abaqus) under a uniformly distributed load $q_0=700$ N/m (see Fig.~6 for reference). The 1-D micropolar beam theory takes into account only the global behavior of lattice core beams.}
\end{figure}
\section{Concluding remarks}
In this paper, a displacement-based geometrically nonlinear finite element model for a micropolar Timoshenko beam which can undergo moderate rotations was developed. Appropriate reduced integration techniques were used to prevent the shear and membrane locking of the elements. The beam was used as an equivalent single layer model for lattice core sandwich beams. Constitutive equations were derived for four different lattice cores (hexagonal, corrugated, Y-frame and web-core) using a two-scale energy approach. The global bending results obtained from the developed 1-D finite element model were in good agreement with 2-D beam frame finite element results calculated using the finite element software ABAQUS.

Although the developed model predicts the global deflections with good accuracy, it does not capture the local buckling of stretch-dominated lattice cores. Nevertheless, the model may be extended to local buckling by allowing stiffness reduction inside the microscale unit cells within a computational multiscale finite element framework. Such considerations, and micropolar plates, are left for future studies.
\section*{Acknowledgements}
The second author acknowledges that this work has received funding from the European Union's Horizon 2020 research and innovation programme under the Marie Sk\l{}odowska--Curie grant agreement No 745770. The financial support is greatly appreciated. The authors also wish to acknowledge CSC -- IT Center for Science, Finland, for computational resources (Abaqus usage).
\appendix
\section{Transformation matrices}
The displacement and strain transformation matrices in Eq.~(8) are
\begin{equation}
\mathbf{T}^{c}_u=
\left[
\begin{array}{cccccccccccc}
 1 & 0 & 0 & 1 & 0 & 0 & 1 & 0 & 0 & 1 & 0 & 0 \\
 0 & 1 & 0 & 0 & 1 & 0 & 0 & 1 & 0 & 0 & 1 & 0 \\
 0 & 0 & 0 & 0 & 0 & 0 & 0 & 0 & 0 & 0 & 0 & 0 \\
 \frac{h}{2} & -\frac{l}{2} & 1 & \frac{h}{2} & \frac{l}{2} & 1 & -\frac{h}{2} & \frac{l}{2} & 1 & -\frac{h}{2} & -\frac{l}{2} &
   1 \\
\end{array}
\right]^{\textrm{T}}
\end{equation}
and
\begin{equation}
\mathbf{T}^{c}_\varepsilon=
\left[
\begin{array}{cccccccccccc}
 -\frac{l}{2} & 0 & 0 & \frac{l}{2} & 0 & 0 & \frac{l}{2} & 0 & 0 & -\frac{l}{2} & 0 & 0 \\
 \frac{h l}{4} & 0 & 0 & -\frac{h l}{4} & 0 & 0 & \frac{h l}{4} & 0 & 0 & -\frac{h l}{4} & 0 & 0 \\
 -\frac{h}{4} & -\frac{l}{4} & 0 & -\frac{h}{4} & \frac{l}{4} & 0 & \frac{h}{4} & \frac{l}{4} & 0 & \frac{h}{4} & -\frac{l}{4} &
   0 \\
 \frac{h}{4} & -\frac{l}{4} & 0 & \frac{h}{4} & \frac{l}{4} & 0 & -\frac{h}{4} & \frac{l}{4} & 0 & -\frac{h}{4} & -\frac{l}{4} &
   0 \\
 0 & 0 & -\frac{l}{2} & 0 & 0 & \frac{l}{2} & 0 & 0 & \frac{l}{2} & 0 & 0 & -\frac{l}{2} \\
\end{array}
\right]^{\textrm{T}}
\end{equation}
respectively.
\section*{References}
\bibliographystyle{elsarticle-harv}
\bibliography{microliteraFE}

%% else use the following coding to input the bibitems directly in the
%% TeX file.

%\begin{thebibliography}{00}

%% \bibitem[Author(year)]{label}
%% Text of biblioBy using the general solution the stress distributions along the beam are presented in terms of the load resultants. graphic item

%\bibitem[ ()]{}

%\begin{align}
%U_x(x\pm l/2,\pm h/2)&= u_x\pm\frac{h}{2}\phi\pm\frac{l}{2}\left(u_x'\pm\frac{h}{2}\phi'\right) , \\
%U_y(x\pm l/2,\pm h/2)&=u_y\pm\frac{l}{2} u_y' , \\
%\Psi(x\pm l/2,\pm h/2)&=\psi\pm\frac{l}{2} \psi' .
%\end{align}

%\end{thebibliography}
\end{document}